\documentclass[review]{elsarticle}

\usepackage{hyperref}

\begin{document}

\begin{frontmatter}

\title{Optimization of One-parameter Family of Integration Formulae
for Solving Stiff Chemical-kinetic ODEs}


\author[mymainaddress]{Youhi Morii}
\cortext[mycorrespondingauthor]{Corresponding author}
\ead{morii@ifs.tohoku.ac.jp}
\address[mymainaddress]{Institute of Fluid Science, Tohoku University, 2-1-1 Katahira, Aoba-ku, Sendai 980-8577, Japan.}

\author[mysecondaryaddress]{Eiji Shima}
\address[mysecondaryaddress]{Research and Development Directorate, Japan Aerospace Exploration Agency, 3-1-1 Yoshinodai, Chuo-ku, Sagamihara, Kanagawa 252-5210, Japan.}

\begin{abstract}
A fast and robust Jacobian-free time-integration method---called Minimum-error Adaptation of a Chemical-Kinetic ODE Solver (MACKS)---for solving stiff ODEs pertaining to chemical-kinetics is proposed herein.
The MACKS formulation is based on optimization of the one-parameter family of integration formulae coupled with a dual time-stepping method to facilitate error minimization.
The proposed method demonstrates higher accuracy compared to the method---Extended Robustness-enhanced numerical algorithm (ERENA)---previously proposed by the authors. Additionally, when this method is employed in homogeneous-ignition simulations, it facilitates realization of faster performance compared to the Variable-coefficient ODE solver (VODE).
\end{abstract}


\end{frontmatter}


\section{Introduction}
Computational fluid dynamic (CFD) analysis constitutes an integral part of and plays a significant role in practical engine development for automotive applications. 
However, from the viewpoint of enhancing engine efficiency and meeting ever-so-stringent emission regulations imposed in recent years, conventional combustion models such as flamelet and G-equation models\cite{Veynante2002} are considered to be insufficient.
Use of a  detailed chemical-reaction model, capable of accurately simulating ignition and extinction phenomena, therefore, becomes necessary.

Currently, zero- and one-dimensional simulations are primarily performed using detailed chemical-kinetics models using Chemkin-Pro and Cantera\cite{Goodwin2018}.
These simulations, however, only provide limited information with regard to the development of actual engine combustors.
Performing CFD analysis using detailed chemical-kinetic models, therefore, becomes necessary.
The difficulty encountered while performing CFD analysis using detailed chemical-kinetics models stems from the associated increase in the number of advection equations, diffusion-coefficient calculations, and stiffness of chemical-reaction ordinary differential equations (ODEs).
However, extant studies have reported that the primary limiting factor in this regard corresponds to solving stiff chemical-kinetic ODEs, and that use of efficient integration methods can help in reducing the computational cost involved therein\cite{Lu2009, Terashima}.

Two main approaches have previously been considered to develop highly efficient integration methods for solving chemical-kinetic ODEs.
The first involves efficient use of implicit methods conventionally employed for solving stiff ODEs\cite{Brown1989,Keken1995,Hairer1988}.
However, as the number of chemical species increases, the computational cost associated with the treatment of Jacobian matrices increases from being proportional to the square of the number of chemical species to the cube of the same number. Additionally, the said Jacobian matrix becomes sparse at the same time\cite{Mott2000,Oran2005,Gou2010,Morii2016}.
The possibility of efficient reduction in computational cost via application of the sparse-matrix method has previously been reported in several studies. Of the various methods prescribed in this regard, the SpeedCHEM\cite{Perini2014} approach is famous and has been widely employed.
In addition, Wang et al. proposed use of a species-clustered splitting scheme during integration\cite{Wang2019}.
The above approaches exponentially reduce the computational cost, thereby facilitating efficient treatment of the Jacobian matrix.

The second approach for solving chemical-kinetic ODEs involves application of robust explicit methods\cite{Mott2000,Gou2010,Morii2016,Qureshi2007,Terashima2015}.
Owing to use of the Courant--Friedrich--Lewy (CFL) number in compressible CFD analysis, the time-step size can be limited to very small values---of the order of $1\times 10^{-8}$ s (for example) or less.
This consideration alone warrants the applicability of robust explicit methods for obtaining reliable solutions to said stiff ODEs.
Since the use of explicit methods is not based on the treatment of Jacobian matrices, a significant reduction in computational cost thereof can be made possible. The
CHEMEQ2\cite{Brown1989}, multi-time-scale method (MTS)\cite{Gou2010}, and ERENA\cite{Morii2016} are examples of recently developed explicit methods widely employed in CFD analysis.

The proposed study focuses on development of a Jacobian-free implicit method to leverage the advantage associated with reduction in computational time involved in Jacobian generation. This is all-the-more true with regard to cases involving sparse Jacobians.
The primary objective of this research concerned the development of a fast and accurate integration method for solving stiff chemical-kinetic ODEs.
Formulation of the proposed method has first been explained in this paper. The said formulation was based on optimization of the one-parameter family of integration equations (to facilitate error minimization) coupled with a dual time-stepping method.
Lastly, performance of the proposed integration method was investigated via comparison of results obtained upon its use during ignition simulations against those obtained using the ERENA (proposed previously by the authors) and conventional VODE\cite{Brown1989} methods.
\section{Methodology}
\subsection{Chemical-kinetic ODEs and model equation}
Chemical-kinetic ODEs under constant-volume adiabatic conditions can be expressed as
\begin{eqnarray}
        \frac{\mathrm{d}y_i}{\mathrm{d}t}=\frac{\dot{\omega}_i}{\rho} \equiv c_i-\frac{y_i}{\tau_i} \ \ \ (i=1,\cdots,N_s) 
        \label{Eq1},
\end{eqnarray}
where $y$ denotes the mass fraction; subscript $i$ denotes the $i$th chemical species; $\dot{\omega}$ denotes the production rate; $\rho$ denotes density; $\tau$ denotes the chemical characteristic time; $c$ denotes the creation rate; and $N_s$ is the total number of species.
A model equation was considered in this study to simplify the discussion concerning proposition of a new integration method for solving chemical-kinetic ODEs.
For construction of the said model equation, the quasi-steady-state (QSS) assumption\cite{Mott2000,Morii2016} was applied to Eq. (\ref{Eq1}); i.e., parameters $c_i$ and $\tau_i$ were assumed to have constant values in Eq. (\ref{Eq1}).
Equation (\ref{Eq1}) in terms of variable $u$ with constants $\alpha$ and $\beta$ can, thence, be expressed as
\begin{eqnarray}
        \frac{\mathrm{d}u}{\mathrm{d}t}=\alpha - \frac{u}{\beta}.
        \label{Eq2}
\end{eqnarray}

The following passages describe the development of the proposed integration method based on the modified model equation Eq. (\ref{Eq2}).

\subsection{Dual Time-stepping Method (DTS)}
To investigate characteristics of the above model equation, the exact solution to Eq. (\ref{Eq2}) can be expressed as 
\begin{eqnarray}
        u=u_0 \exp\left(-\frac{t}{\beta}\right)+\alpha\beta \left[1-\exp\left(-\frac{t}{\beta}\right)\right],
        \label{Eq3}
\end{eqnarray}
wherein $u_0$ represents the initial value of $u$.
When $t\rightarrow \infty$, $u\rightarrow \alpha\beta$; i.e., the product $\alpha\beta$ is representative of a converged solution for $u$.

Next, Eq. (\ref{Eq2}) can be discretized as follows using the explicit Euler (EE) method.
\begin{eqnarray}
        u^{n+1}=u^n+h \left(\alpha-\frac{u^n}{\beta}\right),
        \label{Eq4}
\end{eqnarray}
Here, $n$ denotes the current step number.
From the viewpoint of stability analysis and positive mass fraction, $h\leq \beta$.
When $h = \beta$, Eq. (\ref{Eq4}) takes the form
\begin{eqnarray}
        u^{n+1}=u^n+\beta \left(\alpha-u^n/\beta\right)=\alpha\beta,
        \label{Eq5}
\end{eqnarray}
which is identical to the exact converged solution. This implies that a converged solution can be obtained by means of a single-step calculation when employing the EE technique.
Nonlinear chemical-kinetic ODEs are expected to demonstrate the best convergence properties when $h \approx \tau$.
It is to be noted that when adopting the EE approach for solving chemical-kinetic ODEs, the actual time-step size depends on the minimum characteristic time, thereby implying the time-step size to be invariably small.
To overcome this limitation with regard to time-step size, the DTS was employed in this study.

Application of the DTS method facilitates easy attainment of the converged solution $u^{n+1}$.
From Eq. (\ref{Eq2}), the parameter $F(u)$ can be expressed as
\begin{eqnarray}
  F(u)\equiv \frac{\mathrm{d}u}{\mathrm{d}t}-\left(\alpha-\frac{u}{\beta}\right)=0.
        \label{Eq6}
\end{eqnarray}
Using the above result, the converged solution for $F(u)=0$ can be expressed as $u^{n+1}$.
To obtain $F(u)=0$, a steady-state solution to the following ODE must be obtained.
\begin{eqnarray}
  \frac{\mathrm{d}u}{\mathrm{d}t'}=F(u),
        \label{Eq7}
\end{eqnarray}
where $t'$ denotes the pseudo time.
Any value of the pseudo-time-step size $h'$ can be selected to facilitate attainment of rapid convergence, whereas that of $h$ influences time accuracy of the integration method.
Therefore, different values of $h'$ can be chosen corresponding to different chemical species.
It can, therefore, be realized that use of the DTS approach eliminates the dependency on the minimum characteristic time.
Rewriting the expression for $F(u)$ as
\begin{eqnarray}
        \frac{\mathrm{d}u}{\mathrm{d}t'}=\alpha'-\frac{u}{\beta'},
        \label{Eq8}
\end{eqnarray}
the corresponding solution could be obtained using $h' = \beta'$ in the single-step EE calculation.
Consequently, the following subsection focuses exclusively on the development of an efficient solver for $F(u)$.

\subsection{One-parameter Family of Integration Formulae}
\subsubsection{Characteristic of one-parameter family of integration formulae for stiff ODEs\cite{Liniger1969}}
Equation (\ref{Eq6}) discretized using the one-parameter family of integration formulae can be expressed as
\begin{eqnarray}
  F(u) = \frac{u^{n+1}-u^n}{h} - \left[(1-\theta)\dot{u}^{n+1}+\theta \dot{u}^n\right] = 0,
        \label{Eq9}
\end{eqnarray}
where $\theta$ values of 0, 0.5, and 1 imply use of the implicit Euler (IE), trapezoidal rule (TR), and EE methods, respectively.
The one-parameter family of integration formulae is considered to be A-stable if $\theta \leq 1/2$, and if $\theta = 1/2$ (i.e., when employing the TR-based method), Eq. (\ref{Eq9}) becomes one with 2nd order, thereby demonstrating incurrence of the least error.
The TR-rule based ($\theta =0.5$) approach is, therefore, typically employed in this regard.

\begin{figure}[htbp]
\begin{center}
  \includegraphics[width=0.5\textwidth]{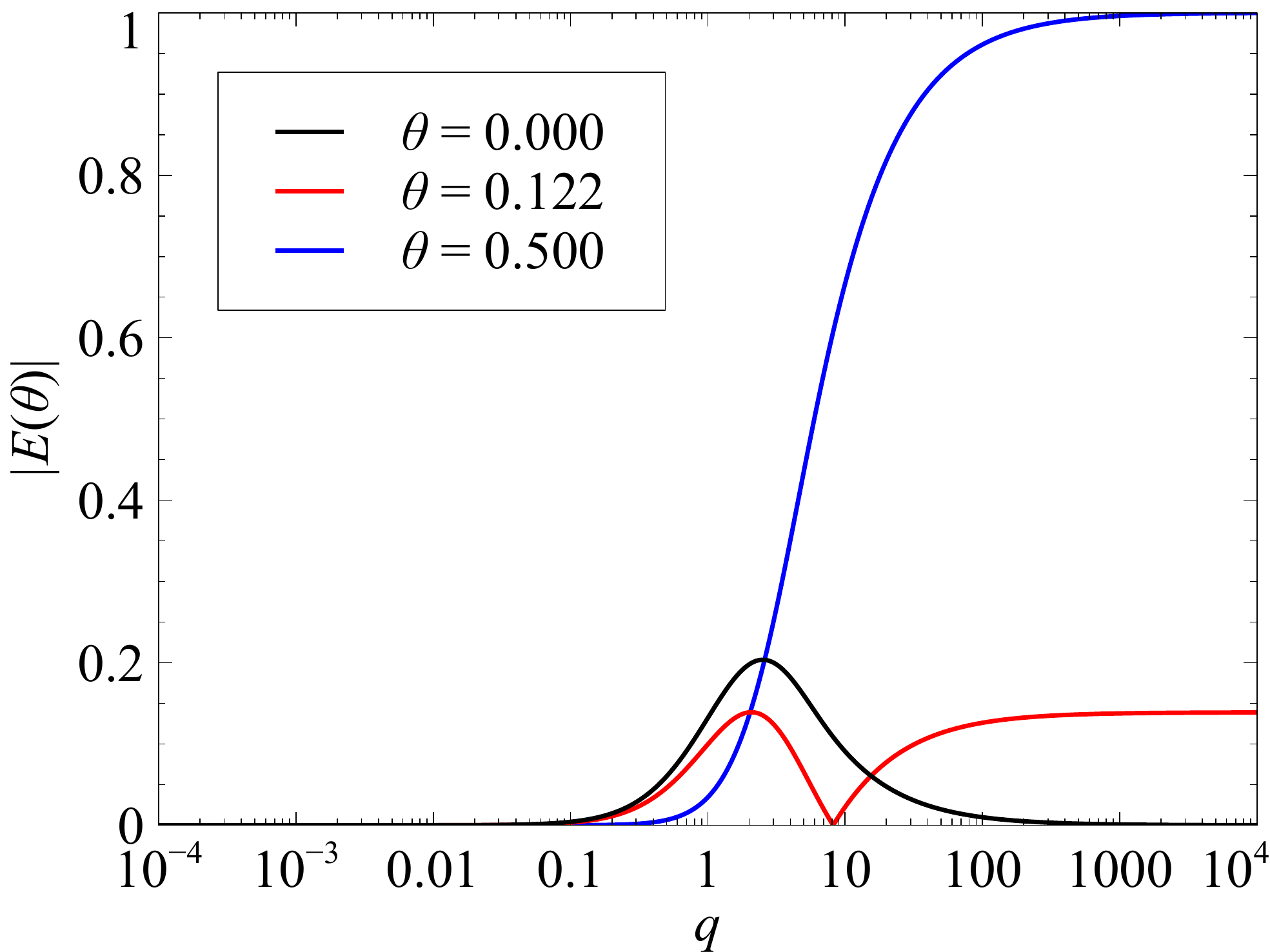} 
  \caption{Relation of $q$ and $E(\theta)$.}
  \label{fig:fig1}
\end{center}
\end{figure}

Using eigenvalues $\lambda$ of Jacobian matrix, the simple stiff system can be expressed as .
\begin{eqnarray}
        \dot{u} = -\lambda u. 
        \label{Eq10}
\end{eqnarray}
With reference to Eq. (\ref{Eq9}), $u^{n+1}$ can be expressed as
\begin{eqnarray}
  u^{n+1} = \frac{1-\theta q}{1+(1-\theta)q}u^{n},
        \label{Eq11}
\end{eqnarray}
where $q = \lambda h$.
The exact solution to $\dot{u} = -\lambda u$ satisfies the relation
\begin{equation}
  u(t^{n+1}) = e^{-q}u(t^n).
        \label{Eq12}
\end{equation}
Using Eqs.(\ref{Eq11}) and (\ref{Eq12}), the relation
\begin{equation}
  \frac{1-\theta q}{1+(1-\theta)q} \approx e^{-q}
        \label{Eq13}
\end{equation}
can be obtained.
Equation (\ref{Eq13}) holds for cases wherein $q \ll 1$.
However, when $q \gg 1$ (i.e., when ODEs are very stiff), Eq. (\ref{Eq13}) is no longer applicable because
\begin{eqnarray}
 \lim_{q \rightarrow \infty} e^{-q}= 1 \ \mathrm{and}  \ 
 \lim_{q \rightarrow \infty} \frac{1-\theta q}{1+(1-\theta)q} = -1.
\end{eqnarray}
The TR-based approach is, therefore, inaccurate when handling stiff ODEs ($q\gg 1$).
To minimize the error incurred when handling stiff ODEs, the value of $\theta$ must be optimized.
In accordance with Liniger's definition, the error induced based on the value of $\theta$ can be expressed as
\begin{eqnarray}
  E(\theta) = \max_{0\leq q \leq \infty} \left| \frac{1-\theta q}{1+(1-\theta)q} - e^{-q} \right|.
  \label{Eq15}
\end{eqnarray}
Based on the above equation, Liniger demonstrated that a value of $\theta = 0.122$ is optimum with regard to minimizing the total error incurred.
Figure \ref{fig:fig1} graphically demonstrates the relation between parameters $q$ and $E(\theta)$ predicted via application of the IE, TR, and Liniger's (optimized one-parameter integration) approaches.
Reference to the said figure demonstrates that the total error incurred when using the TR-based approach increases with increase in ODE stiffness, and that the EI-based method demonstrates comparatively higher accuracy when solving stiff ODEs.
Additionally, as can be observed in the figure, the maximum error incurred when employing Liniger's optimized one-parameter integration approach is the smallest.
It must, however, be noted that ODEs employed in numerical modeling of combustion simulations, in general, demonstrate stiffness.
It can, therefore, be inferred that when attempting to solve stiff chemical-kinetic ODEs, the IE-based approach must be employed to facilitate error reduction. In contrast, the TR-based approach must be exclusively employed under non-stiff conditions.

\subsection{One-parameter Family of Integration Formulae to solve stiffness chemical-kinetic ODEs}
\begin{figure}[htbp]
\begin{center}
  \includegraphics[width=0.5\textwidth]{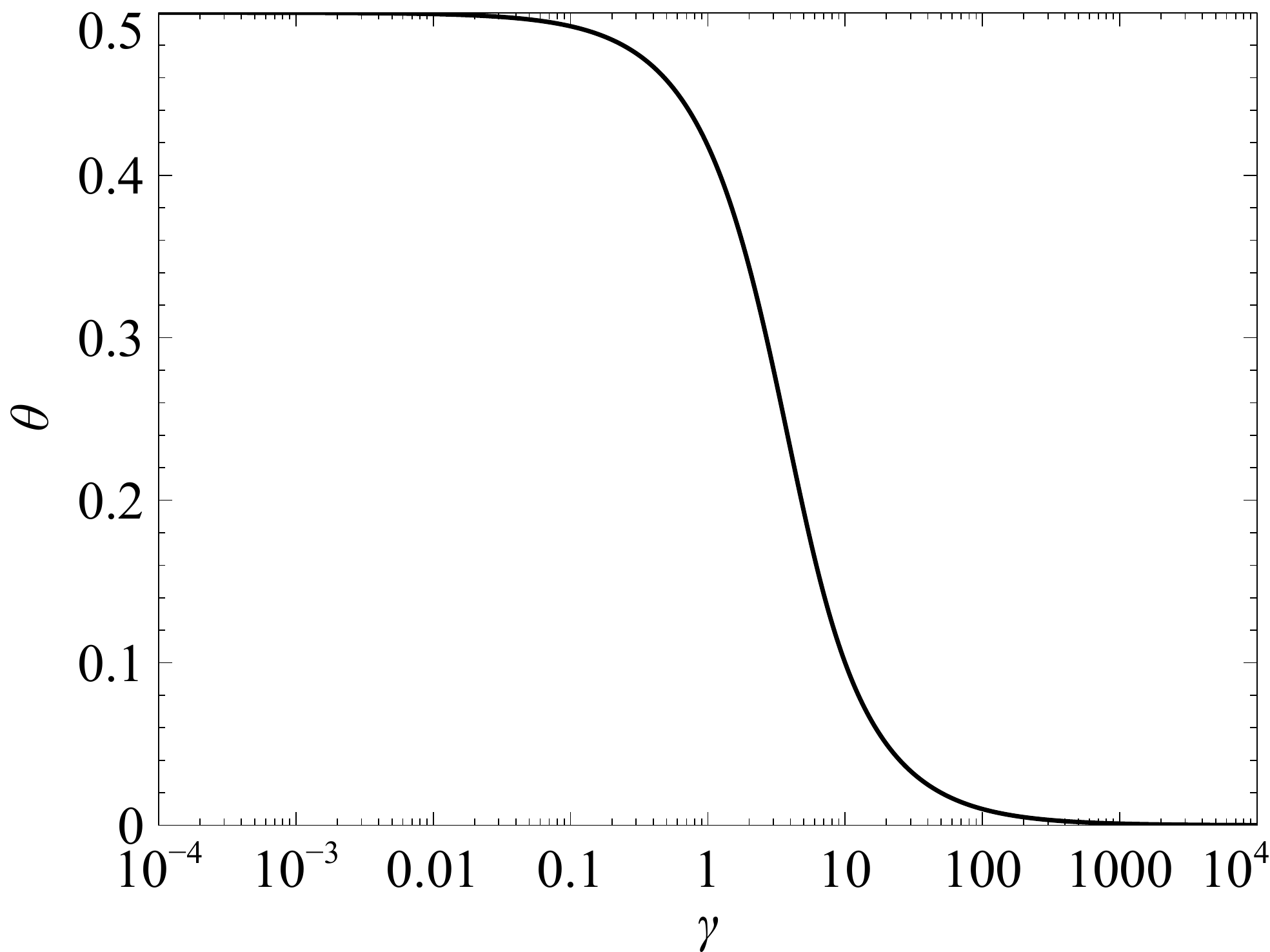} 
  \caption{Relation of $\gamma$ and $\theta$.}
  \label{fig:fig2}
\end{center}
\end{figure}

When Eq. (\ref{Eq5}) is applied to Eq. (\ref{Eq15}), the term which is not related to the exponential could not be deleted.
Consequently, Eq. (\ref{Eq5}) was differentiated in this study.
The resulting equation can be expressed as
\begin{eqnarray}
  \frac{\mathrm{d}u}{\mathrm{d}t} = \left(\alpha-\frac{u^n}{\beta}\right)\exp\left(-\frac{t}{\beta}\right).
  \label{Eq16}
\end{eqnarray}
In addition, the new error estimation has to be applied to discuss the optimization because Eq. (\ref{Eq16}) can not be applied to Liniger's error definition.
Therefore, we defined the new error estimation as
\begin{eqnarray}
  |E|=\left|\left((1-\theta) \frac{\mathrm{d}u}{\mathrm{d}t}^{n+1}+\theta\frac{\mathrm{d}u}{\mathrm{d}t}^{n}\right)-\int^{t^{n+1}}_{t^n}\frac{\mathrm{d}u}{\mathrm{d}t}\mathrm{d}t\right|.
  \label{Eq17}
\end{eqnarray}
Finally, using Eqs. (\ref{Eq16}) and (\ref{Eq17}) along with the assumptions of $t^n = 0$ and $t^{n+1} = h$, $\theta$ could be expressed using $|E| = 0$ as
\begin{eqnarray}
  \theta = \frac{(1-e^{-\gamma})-\gamma e^{-\gamma}}{(1-e^{-\gamma})\gamma}.
\end{eqnarray}
The trend concerning variation in the value of $\theta$ corresponding to changes in $\gamma$ is depicted in Fig. \ref{fig:fig2}.
As can be seen in the figure, the value of $\theta$ can be optimized in accordance with the stiffness of the model equation.
This implies that in the proposed approach, the TR-based method is automatically applied when $\lambda \ll 1$, and the same is true with regard to the IE method when $\lambda \gg 1$.
Consequently, the proposed study's objective of developing a numerical method based on the said model equation can be considered accomplished.

\subsection{Minimum-error Adaptation of Chemical-kinetic ODEs Solver (MACKS)}
The proposed method based on the optimized value of $\theta$ and coupled with the DTS approach has been named as "Minimum-error Adaptation of Chemical-kinetic ODEs Solver (MACKS)."
The corresponding model equations can be expressed as 
\begin{eqnarray}
        \frac{\mathrm{d}u}{\mathrm{d}t'}=\alpha'+\beta' u
\end{eqnarray}
where
\begin{eqnarray}
        \alpha' = \alpha-\theta \frac{u^n}{\beta}+\frac{u^n}{h}, \\
\end{eqnarray}
and
\begin{eqnarray}
        \beta' = \frac{\beta h}{(1-\theta)h + \beta}.
\end{eqnarray}

The MACKS approach when employed for solving chemical-kinetic ODEs can be summarized in the inner-iteration form as
\begin{eqnarray}
y_i^{k+1}=y_i^k+h'_i \left[(1-\theta_i)\dot{\omega}_i^k+\theta_i\dot{\omega}_i^n-\frac{y_i^k-y_i^n}{h}\right],
\end{eqnarray}
where
\begin{eqnarray}
        \theta_i = \frac{1-\exp(-\frac{h}{\tau_i})+\frac{h}{\tau_i}\exp(-\frac{h}{\tau_i})}{\frac{h}{\tau_i}(1-\exp(-\frac{h}{\tau_i}))},
\end{eqnarray}
Additionally, as described above, when $h' = \beta '$, a converged solution can be obtained. The corresponding relation for evaluating $h'$ can be expressed as
\begin{eqnarray}
h' = \frac{\tau_i h}{(1-\theta_i)h + \tau_i}.
\end{eqnarray}

The proposed MACKS approach employs the absolute- and relative-error thresholds (ATOL and RTOL, respectively) to facilitate error evaluation in a manner similar to the VODE approach.
To preserve estimation accuracy, the actual time-step size $h$ was divided by 4 for cases wherein the evaluated error increased within the inner loop.

\section{Results and Discussions}
\subsection{Numerical conditions}
This study compares the proposed MACKS approach against previously proposed VODE and ERENA methods in terms of their prediction accuracy and computational cost with regard to solving several zero-dimensional, homogeneous ignition problems involving stoichiometric mixtures of H$_2$/air, CH$_4$/air, $n$-C$_7$H$_{16}$/air, and $n$-C$_{10}$H$_{22}$/air maintained under conditions of $p_0$ = 0.1 and 1.0 MPa and $T_0$ = 1300 K.
Values of RTOL and ATOL corresponding to the MACKS and VODE approaches were observed to be $1\times 10^{-5}$ and $1\times 10^{-13}$, respectively.
The corresponding threshold value for ERENA was set as $1\times 10^{-8}$.
The tolerances and threshold are the same as in our previous study\cite{Morii2016}.
Detailed reaction mechanisms considered in this study were generated in accordance with the Knowledge-basing Utilities for Complex Reaction Systems (KUCRS)\cite{Miyoshi2005}, and details concerning the number of chemical species and reactions considered have been summarized in Table \ref{tab_a_1}.
It must be noted that the Jacobian matrix concerning the VODE approach was intentionally initialized during each time step owing to the requirement of an initialization process when performing multidimensional CFD simulations.
The corresponding base time-step sizes were $h = 1\times 10^{-8}$ and $1\times 10^{-7}$ s.

\subsection{Accuracy and computational cost}
\begin{figure}[htbp]
\begin{center}
  \includegraphics[width=0.5\textwidth]{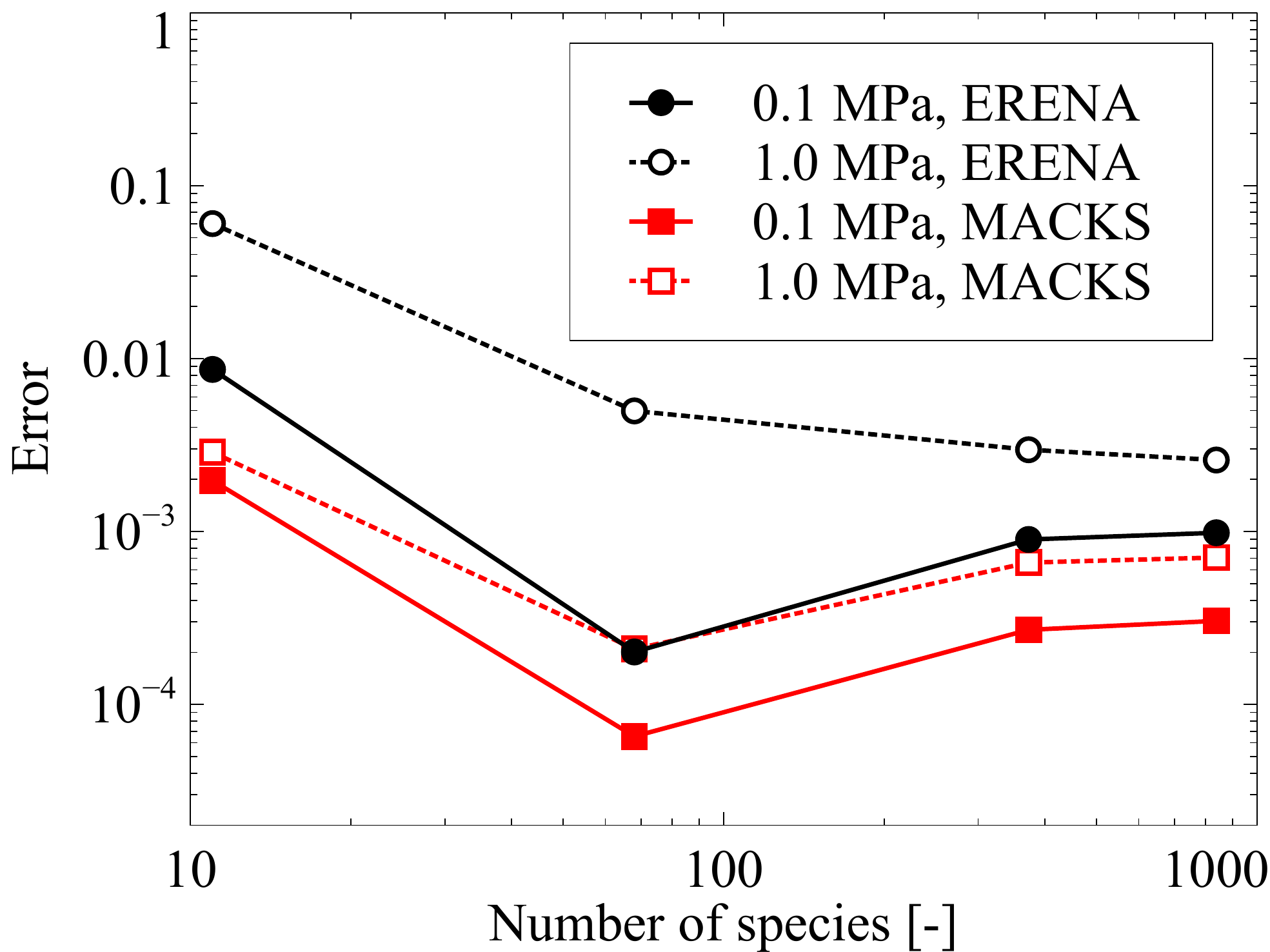} 
  \caption{Ignition delay error of ERENA and MACKS for the H$_2$/air, CH$_4$/air, nC$_7$H$_{16}$ /air, and nC$_{10}$H$_{22}$/air mixtures where $h=1\times 10^{-8}, p_0=0.1 and 1$ MPa, and $T_0=1300$ K.}
  \label{fig:fig3}
\end{center}
\end{figure}

In this study, the accuracy with regard to the ignition-delay time estimated when employing the MACKS approach was first compared against that estimated using the combined VODE and ERENA approaches with $h = 1\times 10^{-8}$ s.
The ignition-delay-time error was defined as
\begin{equation}
        \mathrm{Error}  = \frac{{{\mathrm{IDT}}}_\mathrm{TIM}-{\mathrm{IDT}}_\mathrm{VODE}}{{\mathrm{IDT}}_\mathrm{VODE}},
        \label{eq_c_1}
\end{equation}
where IDT represents the ignition-delay time and subscript TIM corresponds to ERENA or MACKS.
As can be realized, if the value of "Error" is small, results obtained using the proposed approach demonstrate good agreement with those obtained using VODE.
Reference to Fig. \ref{fig:fig3} demonstrates the MACKS approach to be more accurate compared to ERENA.
In particular, when the number of chemical species is small, use of the MACKS approach drastically reduces the error incurred when compared against the ERENA approach.
This is because when employing the ERENA approach, the mass fraction is divided at every time step to conserve the summation of mass fractions \cite{Morii2016} in accordance with the equation
\begin{eqnarray}
y^{n+1}_i = \frac{y^*_i}{\sum_{i = 1}^{N_s} y^*_i}.
\end{eqnarray}
Using the ERENA approach's error threshold $\varepsilon$, the resulting summation of mass fractions can be expressed as
\begin{eqnarray}
1 = \sum_{i=1}^{N_s} y^{n+1}_i = \left(\sum_{i=1}^{N_s}y_i^*\right)  \pm \varepsilon \approx \sum_{i=1}^{N_s}\left(y_i^*  \pm \frac{\varepsilon}{N_s}\right).
\end{eqnarray}
The optimization error associated with the use of the ERENA approach can, therefore, be expressed as
\begin{eqnarray}
y^{n+1}_i \approx y_i^*  \pm \frac{\varepsilon}{N_s}
\end{eqnarray}
Consequently, the maximum error incurred per mass fraction can be represented in terms of $\varepsilon/N_s$.
Thus, the ERENA approach can be considered less accurate when employed in cases involving a small number of species.

Next, the computational cost associated with the MACKS approach was compared against that incurred when employing the VODE and ERENA methods with $h=1\times 10^{-8}$ s.
The ratio of the total computational time could be defined as
\begin{equation}
        \mathrm{Efficient\ ratio} = \frac{({t_\mathrm{CPU}})_\mathrm{VODE}}{({t_\mathrm{CPU}})_\mathrm{TIM}},
        \label{eq_c_2}
\end{equation}
where $t_\mathrm{CPU}$ denotes the total computational time.

\begin{figure}[htbp]
\begin{center}
  \includegraphics[width=0.5\textwidth]{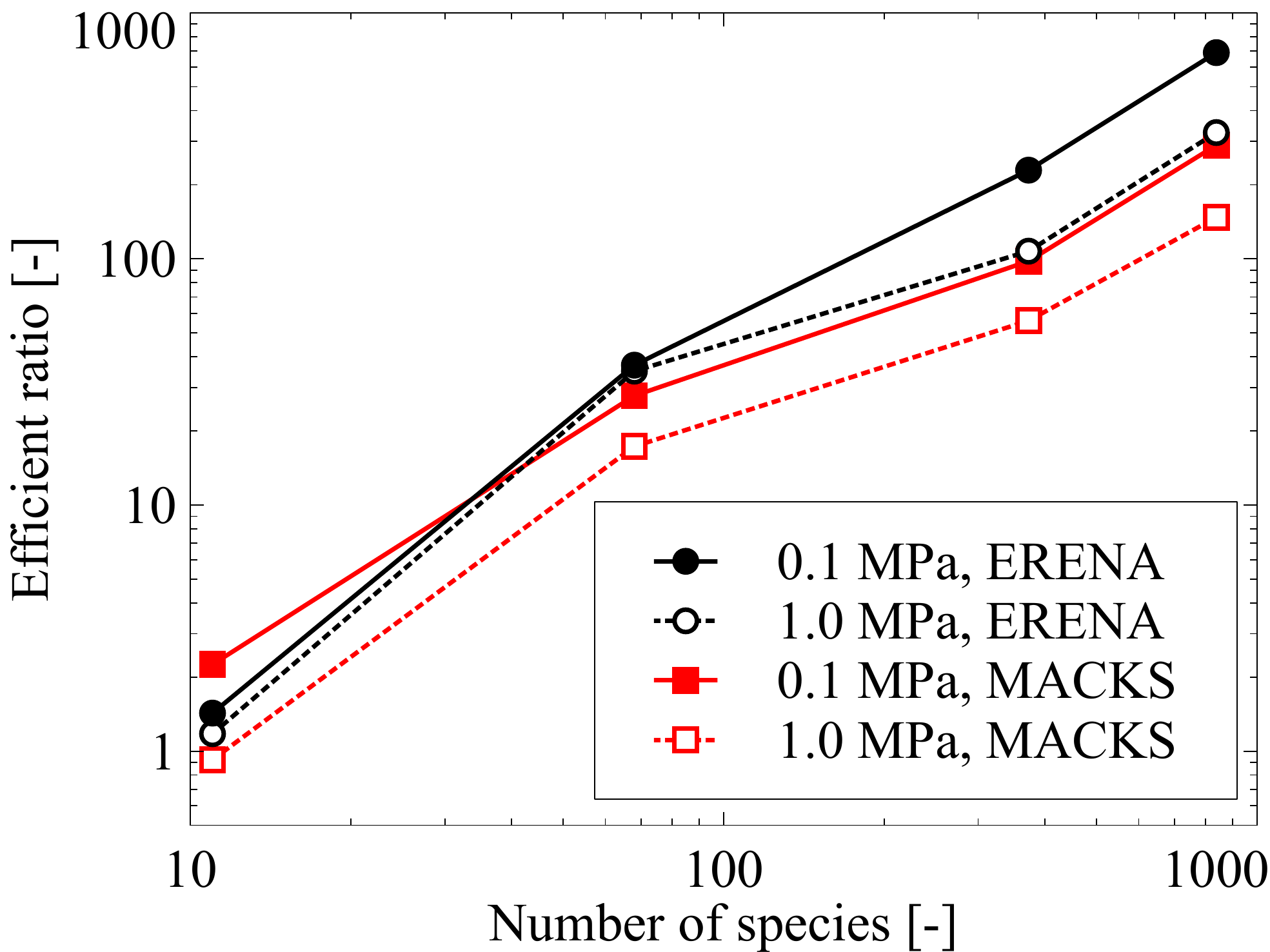} 
  \caption{Total computational cost ratio of ERENA and MACKS for the H$_2$/air, CH$_4$/air, nC$_7$H$_{16}$ /air, and nC$_{10}$H$_{22}$/air mixtures where $h=1\times 10^{-8}, p_0=0.1 and 1$ MPa, and $T_0=1300$ K.}
\end{center}
\label{fig:fig4}
\end{figure}

Reference to the above equation demonstrates that when "Efficient ratio" equals less than 1, the corresponding computational cost exceeds that incurred when using the VODE method; likewise, when "Efficient ratio" exceeds unity, the associated computational cost is lower compared to that incurred using the VODE approach.
Corresponding results are depicted in Fig. \ref{fig:fig4}.
As can be observed, the MACKS approach is slower compared to ERENA for cases wherein the number of chemical species involved is large.
Accordingly, if only a small number of chemical species are involved, performance of the MACKS approach is comparable to that of ERENA.

To facilitate better understanding of characteristics of the MACKS approach, the following subsection discusses the case involving a CH$_4$/air mixture maintained at a pressure of 1 MPa.
This is because chemical-kinetic models, typically, are reduced to involve less than 100 species, since the computational cost incurred becomes impractical when considering more than 100 species.
Additionally, a pressure of 1 MPa is considered to be appropriate with regard to actual applications.

\subsection{Accuracy and computational cost of CH$_4$/air case}
\begin{figure}[htbp]
        \begin{minipage}{0.5\hsize}
                \begin{center}
                        \includegraphics[width=\textwidth]{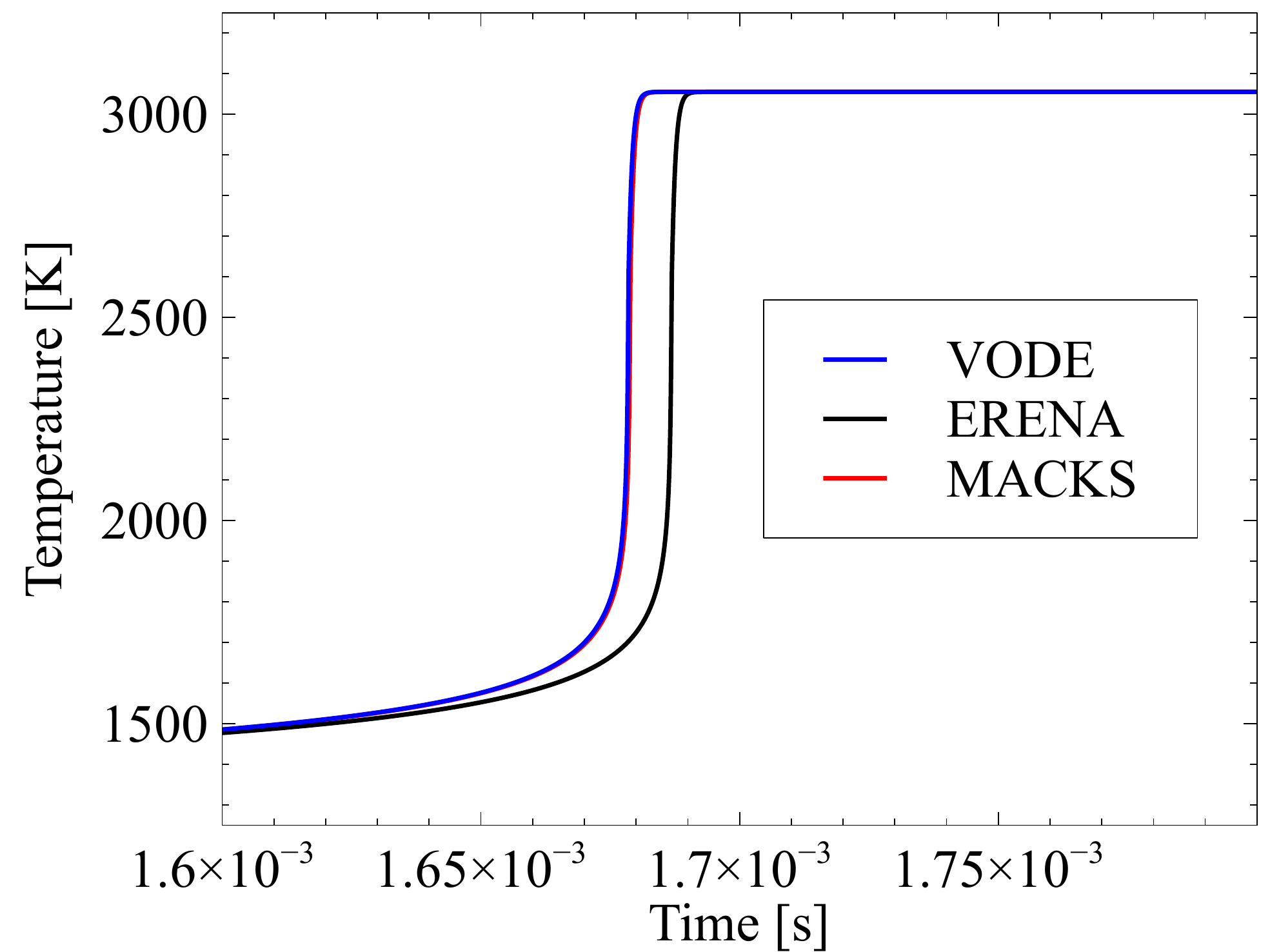}
                        (a) $h = 1\times10^{-8}$ s
                \end{center}
        \end{minipage}
        \begin{minipage}{0.5\hsize}
                \begin{center}
                        \includegraphics[width=\textwidth]{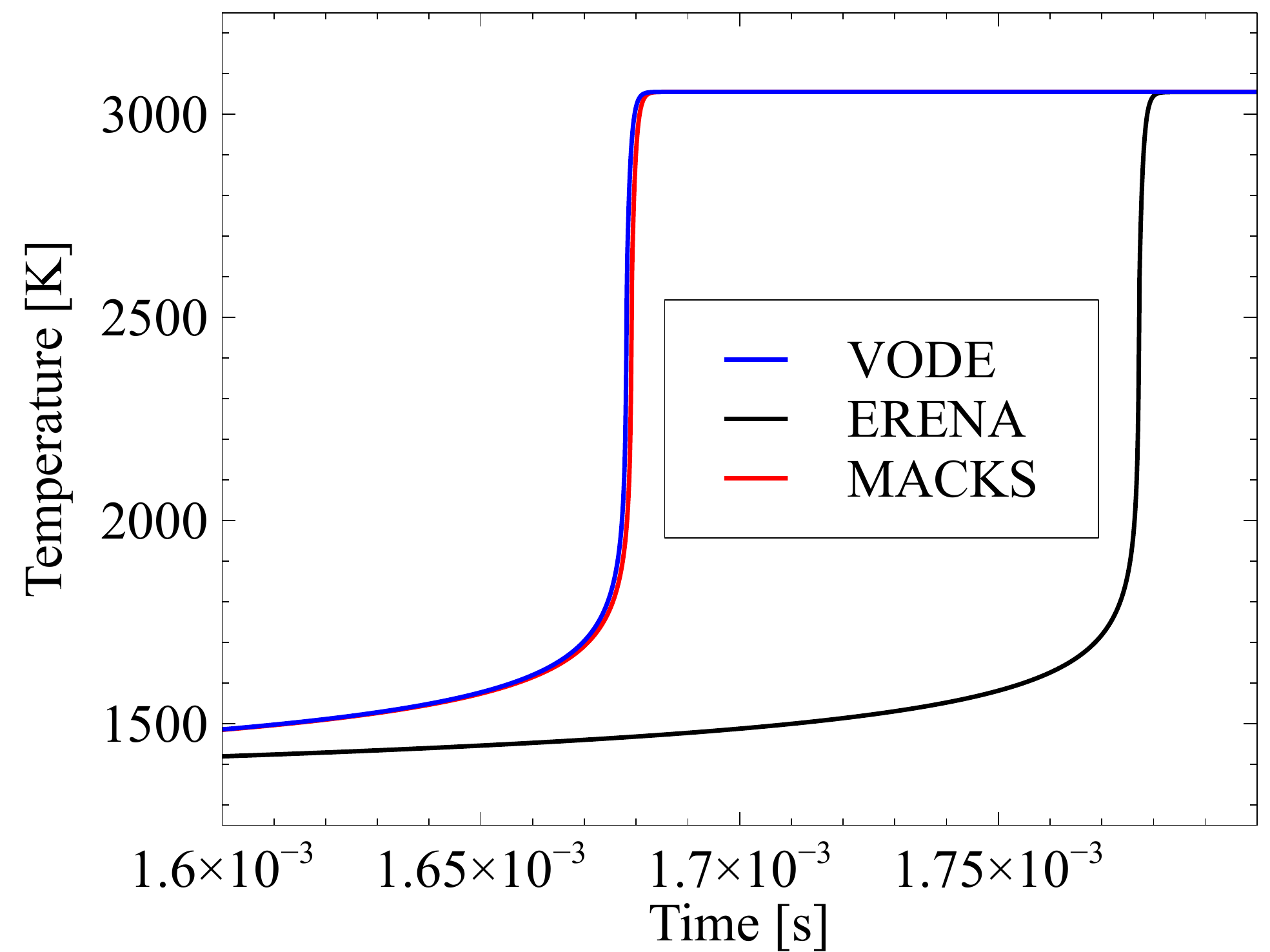}
                        (b) $h = 1\times10^{-7}$ s 
                \end{center}
        \end{minipage}
        \caption{Time histories of temperature with VODE, ERENA, and MACKS for a CH$_4$/air mixture with $h = 1 \times 10^{-8}$ s, $p_0=1.0$ MPa, and $T_0=1300$ K.}
\label{fig:fig5}
\end{figure}

Time histories of temperature variations concerning the CH$_4$/air mixture are depicted in Fig. \ref{fig:fig5} for cases involving $h$ values of $=1\times 10^{-8}$ and $1\times 10^{-7}$ s.
As can be observed, results obtained using the MACKS approach demonstrate good agreement with those obtained using VODE.
However, results obtained using ERENA demonstrate large deviations from those obtained using VODE, especially when $h = 1\times 10^{-7}$ s.

\begin{figure}[htbp]
        \begin{minipage}{0.5\hsize}
                \begin{center}
                        \includegraphics[width=\textwidth]{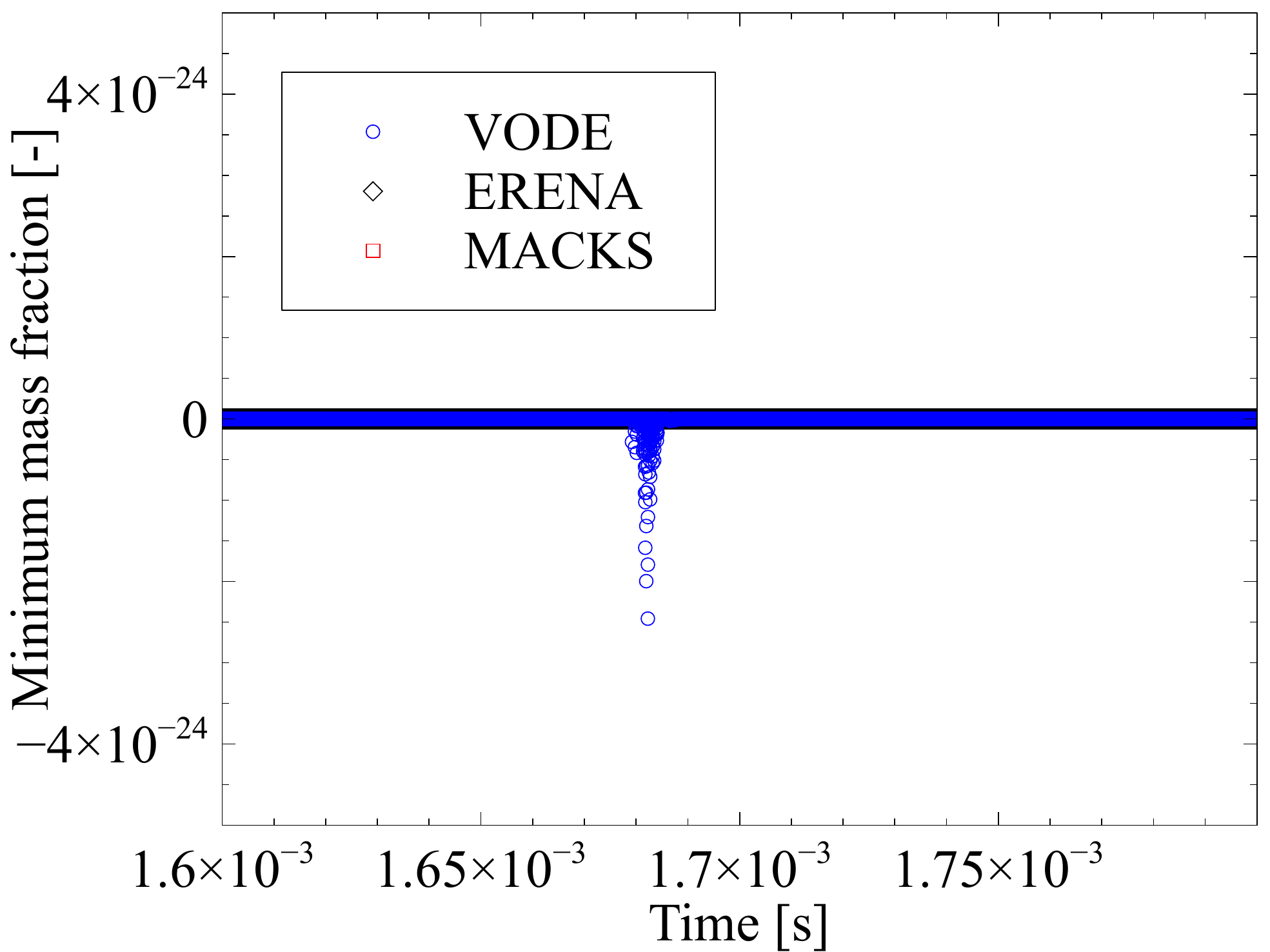}
                        (a) $h = 1\times10^{-8}$ s
                \end{center}
        \end{minipage}
        \begin{minipage}{0.5\hsize}
                \begin{center}
                        \includegraphics[width=\textwidth]{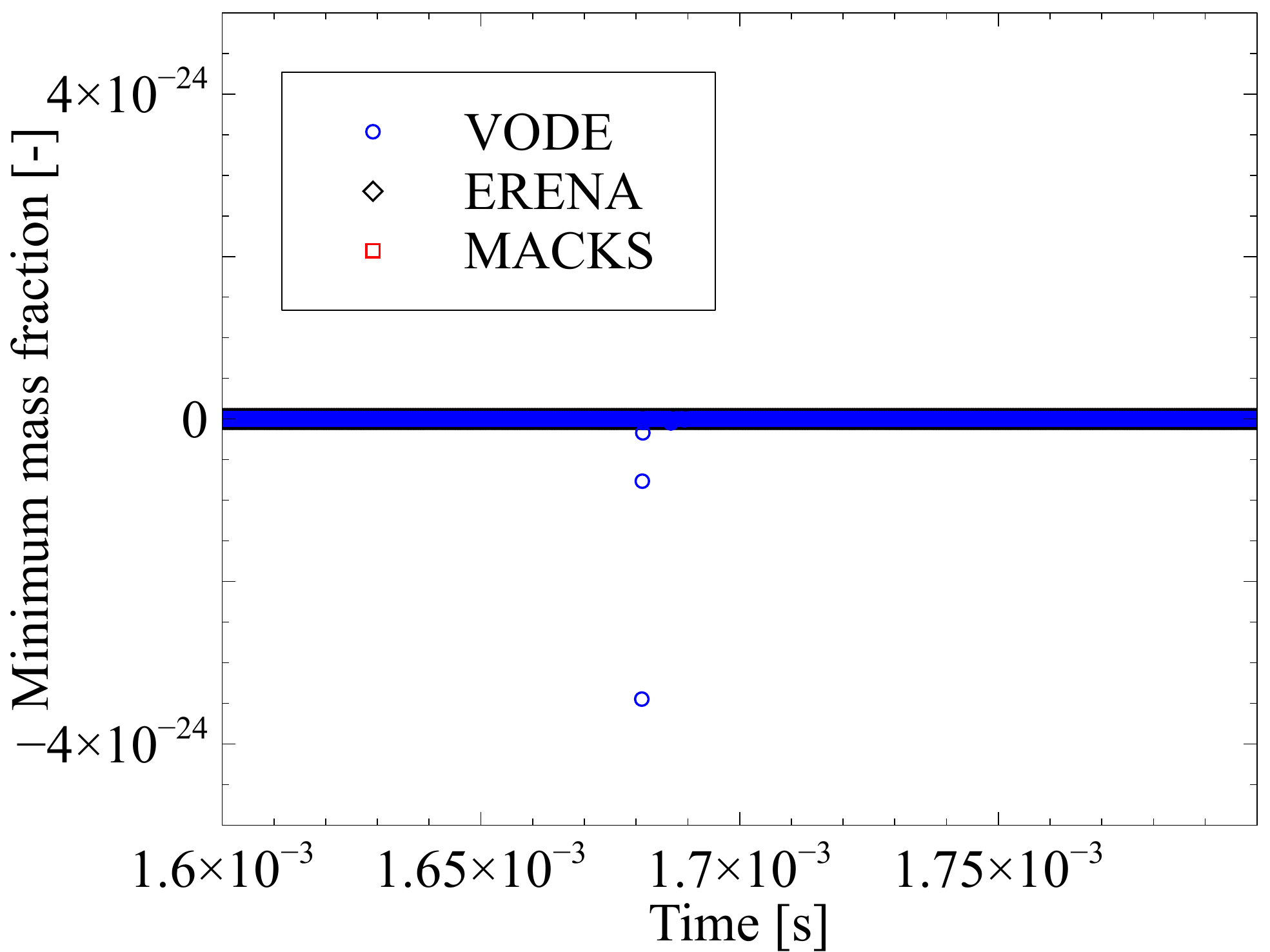}
                        (b) $h = 1\times10^{-7}$ s
                \end{center}
        \end{minipage}
        \caption{Time histories of the minimum mass fraction with VODE, ERENA, and MACKS for a CH$_4$/air mixture with $h = 1 \times 10^{-8}$ s, $p_0=1.0$ MPa, and $T_0=1300$ K.}
\label{fig:fig6}
\end{figure}

Fig. \ref{fig:fig6} depicts time histories of the minimum mass fraction concerning the the CH$_4$/air mixture for cases with $h=1\times 10^{-8}$ and $h=1\times 10^{-7}$ s.
As can be observed, use of the ERENA and MACKS approaches demonstrates zero or positive values of the minimum mass fraction, whereas the VODE mothod demonstrates negative values of corresponding parameters.
This is because both ERENA and MACKS approaches are based on exact solutions of model equations.
Additionally, the VODE approach is well-known for its characteristic of producing negative mass fractions\cite{Nguyen2009}.

\begin{figure}[htbp]
        \begin{minipage}{0.5\hsize}
                \begin{center}
                        \includegraphics[width=\textwidth]{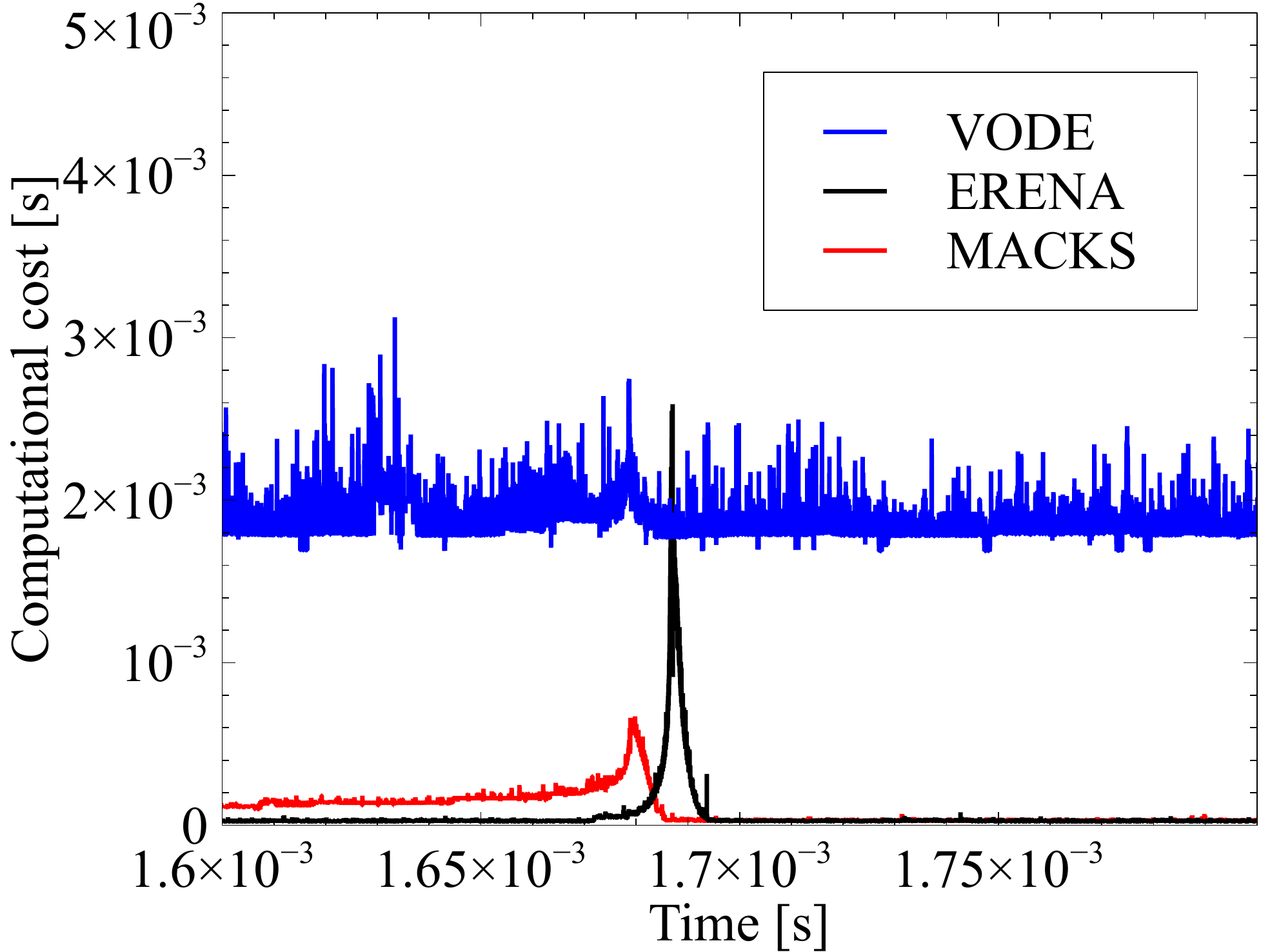}
                        (a) $h = 1\times10^{-8}$ s
                \end{center}
        \end{minipage}
        \begin{minipage}{0.5\hsize}
                \begin{center}
                        \includegraphics[width=\textwidth]{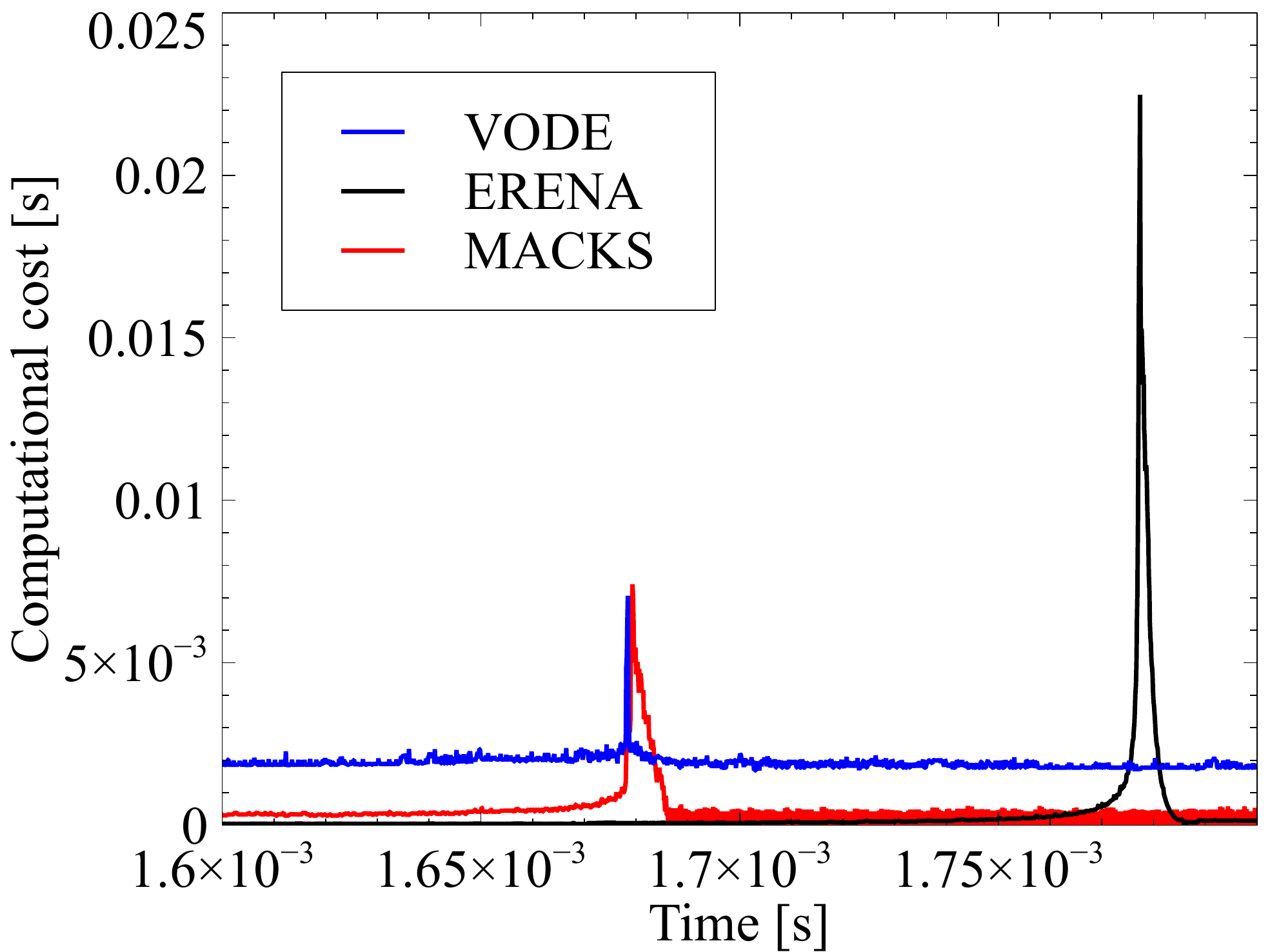}
                        (b) $h = 1\times10^{-7}$ s
                \end{center}
        \end{minipage}
        \caption{Time histories of computational cost per one-iteration with VODE, ERENA, and MACKS for a CH$_4$/air mixture with $h = 1 \times 10^{-8}$ s, $p_0=1.0$ MPa, and $T_0=1300$ K.}
        \label{fig:fig7}
\end{figure}

Lastly, time histories concerning the computational cost incurred per iteration when employing the MACKS, ERENA, and VODE approaches are depicted in Fig. \ref{fig:fig7} for cases with $h=1\times 10^{-8}$ and $h=1\times 10^{-7}$ s.
As can be observed, the peak computational cost corresponding to the MACKS approaches is much smaller compared to those corresponding to the VODE and ERENA approaches with $h=1\times 10^{-8}$ s.
The MPI library is widely employed in high-performance computing, wherein load balancing forms one of the most important considerations.
Figure \ref{fig:fig7} reveals that use of the MACKS approach demonstrates much better load-balancing characteristics compared to the ERENA method.
This is because the difference in computational costs between flame zone and the other is small when the MACKS approach is employed during CFD analysis.
In addition, the peak computational cost associated with the use of the MACKS approach is nearly identical to that associated with VODE when $h = 1\times 10^{-7}$ s.
In other words, the MACKS approach can be expected to demonstrate higher performance compared to the VODE method even with less than 100 chemical species and a large time interval, and to have much higher performance under this condition than ERENA.

\section{Conclusions}
This study proposes a fast and robust Jacobian-free time-integration method called MACKS to facilitate treatment of stiff chemical-kinetic ODEs. The proposed approach was derived using the optimized one-parameter family of integration formulae to facilitate minimization of aggregate error. Results obtained in this study demonstrate that use of the MACKS approach leads to lower computational cost associated with solving zero-dimensional ignition problems whilst achieving higher prediction accuracy compared to the VODE approach.
In addition, the MACKS has been demonstrated to possess higher accuracy compared to the ERENA method previously proposed by the authors.

\bibliography{ODE.bib}
\clearpage

\begin{table} 
  \begin{center}
  \caption{Numbers of species and elementary reactions in detailed reaction mechanisms}
    \label{tab_a_1}
    \vspace{5mm}
    \begin{tabular}[htpb]{ccc}
\noalign{\hrule height 1pt}
      & No. of species & No. of reactions \\
      \hline
      \hline
      H$_2$  & 11 & 34 \\
      CH$_4$  & 68 & 334 \\
      $n$-C$_7$H$_{16}$  & 373 & 1071 \\
      $n$-C$_{10}$H$_{22}$  & 839 & 2126 \\
\noalign{\hrule height 1pt}
    \end{tabular}
  \end{center}

\end{table}

\end{document}